\documentstyle [12pt] {article}

\title{Is the Higgs a visible particle ?
\thanks{\it{This work was partially supported by  Comision de
Investigaciones Cientificas, Pcia. de Buenos Aires; Argentina.}}}
\author{C.G.Bollini, M.C.Rocca\\
Departmento de Fisica, Fac. de Ciencias Exactas,\\
Universidad Nacional de La Plata.\\
C.C. 67 (1900) La Plata. Argentina}
\date{August 1, 1995}

\begin{document}

\maketitle
\begin{abstract}
 We suggest that the Higgs might be unobservable as a free particle,
due to its origin at a symmetry breaking mechanism.

 The standard model is kept intact, only the definition of the
vacuum for the Higgs is changed. With the new
(natural) definition, the Higgs propagator is half advanced and half
retarded. This Green function is compatible with the absence of
free particles.

\end{abstract}

\eject

\section {Introduction}

 The standard model for the electroweak interaction has so far been
succesful in the interpretation of all known experimental results.
Perhaps, the only piece of evidence still lacking is the observation
of the Higgs particle, to which a large mass M has to be attributed.
However, according to references \cite{tp1} \cite{tp2}, for M
$ \geq $ 1 Tev. the standard model becomes "strongly interacting"
and perturbation theory ceases to be valid.

 Nevertheless, we would like to point out that the scalar
$ \phi $ is not a pure Klein-Gordon field. In fact, the lagrangian
potential is quadratic with positive coefficient ($ M^2 $) near
the minimun $ {\phi}_0 $. When $ \phi $ is weak ( $ \mid \phi \mid $
$ \ll $ $ \mid {\phi}_0 \mid $ ) the coefficient of the quadratic
term is negative. Thus signalling, of course, an instability.
But showing that in this case a quantum of $ \phi $ should be
considered to be a tachyon, rather than a bradyon. In view of
the experimental situation, the comparison makes sense, as tachyons
never appear in free particle states (See references \cite{tp3}
\cite{tp4}).

 So, let us admit the possibility that, due to its origin (at a
symmetry breaking mechanism), the Higgs can not manifest itself
as a free particle. What are the consequences of this hypotesis ?.

 The succes of the standard model is an indication that the
equation for the Higgs is basically correct. However we will see
that this equation does not necessarily implies that the quanta
of the field are observable. In fact, without any change in the
model or its quantization, an alternative appears when a definition
of the vacuum has to be adopted. This alternative gives rise
to two different Green functions. One is the usual causal
propagator and the other is a half advanced and half retarded
propagator. The latter is precisely the Green function suitable
for fields that can not reach the asymptotically free states (See
ref.\cite{tp5} and also \cite{tp4} \cite{tp6} \cite{tp7}).
We emphasize again that we are not proposing a change in the model.
We suggest that the Higgs can not ocupy free states. Its
propagator must be related to this condition.

\section{ The states of the Higgs field }

 For the sake of the argument, we are going to repeat here some
well-known properties of a quantized Klein-Gordon field.

 Such a field has the Fourier form:

\begin{equation}
\phi (x) = \frac {1} {(2 \pi )^{3/2}} \int \frac {d^3 k}
{\sqrt{2 {\omega}_{\vec{k}}}} \left( a_{\vec{k}} e^{-ikx} +
 a_{\vec{k}}^{+} e^{ikx}
\right)
\end{equation}

 Where

\begin{equation}
k_0 = {\omega}_{\vec{k}} = {\left( {\vec{k}}^2 + M^2 \right)}^{\frac{1}{2}}
\end{equation}

 The Hamiltonian is:

\begin{equation}
H = \frac {1} {2} \int d^3 k \;\; {\omega}_{\vec{k}}
\left( a_{\vec{k}} a_{\vec{k}}^{+} +
a_{\vec{k}}^{+} a_{\vec{k}} \right)
\end{equation}

 The creation and anihilation operators obey:

\begin{equation}
\left[ a_{\vec{k}}, a_{{\vec{k}}^{'}}^{+} \right] = \delta \left(
 \vec{k} -
{\vec{k}}^{'} \right)
\end{equation}

 For each (discretized) degree of freedom we have:

\begin{equation}
\left[ a, a^{+} \right] = 1
\end{equation}
\begin{equation}
\left[ H, a \right] = - \omega a
\end{equation}
\begin{equation}
\left[ H, a^{+} \right] = + \omega a^{+}
\end{equation}

 Equation (6) (resp.(7)) gives rise to a chain of states with
decreasing (resp. increasing) energies.

 At this point we have the following possibilities:
I) If the quanta of the field correspond to normal physical particles,
the energy can not be negative and the descending ladder must stop
at the so called vacuum state

\begin{equation}
a \mid 0 > = 0
\end{equation}

 It is easy to deduce that

\begin{equation}
< 0 \mid a a^{+} \mid 0 > = 1
\end{equation}
\begin{equation}
< 0 \mid a^{+} a \mid 0 > = 0
\end{equation}

 Or, in the continuum:

\begin{equation}
< 0 \mid a_{\vec{k}} a_{{\vec{k}}^{'}}^{+} \mid 0 > = \delta \left(
 \vec{k} -
{\vec{k}}^{'} \right)
\end{equation}
\begin{equation}
< 0 \mid a_{\vec{k}}^{+} a_{{\vec{k}}^{'}} \mid 0 > = 0
\end{equation}
II) If the field represents particles which never appear as free
entities, then (8) is no longer a necessity and we could have
a never ending descending and ascending chain:

\begin{equation}
a^{+} \mid s > = {\alpha}_s \mid s+1 >
\end{equation}
\begin{equation}
a \mid s > = {\beta}_s \mid s-1 >
\end{equation}
where s is any integer (positive, negative or zero).

 The vacuum is now defined as (cf.(3)):

\begin{equation}
\left( a a^{+} + a^{+} a \right) \mid 0 > = 0
\end{equation}

 This relation (with eq.(5)) imply:

\begin{equation}
a a^{+} \mid 0 > = \frac {1} {2} \mid 0 >
\end{equation}
\begin{equation}
a^{+} a \mid 0 > = - \frac {1} {2} \mid 0 >
\end{equation}

 Leading to the expectation values:

\begin{equation}
< 0 \mid a_{\vec{k}} a_{{\vec{k}}^{'}}^{+} \mid 0 > = \frac {1} {2}
 \delta \left(
\vec{k} - {\vec{k}}^{'} \right)
\end{equation}
\begin{equation}
< 0 \mid a_{{\vec{k}}^{'}}^{+} a_{\vec{k}} \mid 0 > = - \frac {1} {2}
 \delta \left(
\vec{k} - {\vec{k}}^{'} \right)
\end{equation}

 It is easy to realize that case II is a superposition of case I
and its time reversed case for which

\begin{equation}
a_{\vec{k}}^{+} \mid 0 > = 0
\end{equation}
\begin{equation}
< 0 \mid a_{\vec{k}} a_{{\vec{k}}^{'}}^{+} \mid 0 > = 0
\end{equation}
\begin{equation}
< 0 \mid a_{\vec{k}}^{+} a_{{\vec{k}}^{'}} \mid 0 > = - \delta \left(
 \vec{k} -
{\vec{k}}^{'} \right)
\end{equation}

\section{ Propagator for the Higgs }

 The vacuum expectation values of chronological products of field
operators are easily deduced from the relations found in $ \S $ 2.
However, there is no need of an actual calculation.

 It is well-known that case I gives raise to the causal
propagator. It is the Fourier transform of $ (k^2 + M^2)^{-1} $
where tho $ k_0 $ -integration follows Feynman's path:

\begin{equation}
F(k) = \frac {1} {\left(k^2 + M^2 \right)}_F =
\frac {1} {\left( - k_0^2 + {\omega}_{\vec{k}}^2 \right)}_F
\end{equation}

 The subindex is a label to indicate that the $ k_0 $ -integration
goes below the pole at $ k_0 $ = $ - {\omega}_{\vec{k}} $ (advanced),
 and
above the pole at $ k_0 $ = + $ {\omega}_{\vec{k}} $ (retarded).

 Analogously, equation (21) and (22) give raise to the anticausal
Green function.

\begin{equation}
\bar{F} (k) = \frac {1} {\left( k^2 + M^2 \right) }_{\bar{F}} =
\frac {1} {\left( - k_0^2 + {\omega}_{\vec{k}}^2 \right) }_{\bar{F}}
\end{equation}
where the $ k_0 $ -integration goes now above the pole at $ k_0 $
= $ - {\omega}_{\vec{k}} $ and below the pole at
$ k_0 $ = + $ {\omega}_{\vec{k}} $ .

 Clearly, case II is a superposition of (11), (12) and (21), (22).
In other words, the Green function for case II is half causal and
half anticausal. The propagator is the Fourier transform of
$ (k^2 + M^2 )^{-1} $ , but this time the $ k_0 $ -integration
is half advanced and half retarded at both poles.

 This type of Green function ("Wheeler propagator") was used in
ref. \cite{tp5} to describe the electromagnetic interaction of
perfect absorbers. I.e. when no free radiation escapes the
system.

 It is interesting that in spite of the fact that this propagator
contains advanced as well as retarded effects, no violation of
causality can be observed (See ref. \cite{tp8}).

 There is a simple relation between the advanced and the retarded
integration at a pole $ k_0 $ = $ \pm $ $ {\omega}_{\vec{k}} $ .

\begin{equation}
\frac {1} {\left[ k_0 - \omega \right] }_{ad} =
\frac {1} {\left[ k_0 - \omega \right] }_{rt} +
2 \pi i \delta \left( k_0 - \omega \right)
\end{equation}

 Equation (25) is due to the fact that the difference between the
advanced and the retarded path is a positive loop around the pole.
By Cauchy's theorem this loop integration is equivalent to
$ 2 \pi i $ times a $ \delta $ -function.

 As a consequence of (25), there is also a simple relation between
Feynman's and Wheeler's propagators:

\begin{equation}
F(k) = W(k) + i \pi \delta \left( k^2 + M^2 \right)
\end{equation}

 Equation (26) is also valid for complex-mass parameters (See
ref. \cite{tp7} ).

 When $ M^2 $ is real and positive, W(k) can be recognized as
Cauchy's principal value Green function. So that, in this case,
equation (26) is identical to the well-known relation:

\begin{equation}
F(k) = P \frac {1} {k^2 + M^2} + i \pi \delta \left( k^2 + M^2
\right)
\end{equation}

 The Wheeler Green function, for $ M^2 $ $ \geq $ 0 ,

\begin{equation}
W(k) = P \frac {1} {k^2 + M^2} = P \frac {1} {- k_0 +
{\omega}_{\vec{k}}^2 }
\end{equation}
is exactly zero on the mass shell.

 Equation (26) can be considered to be a separation of Feynman's
causal function into two terms. The $ \delta $ -function represents
the free propagation of the particle, while W(k) contains only the
virtual propagation.

 It is also worth mentioning, that case II is the unique superposition
of causal and anticausal propagators that gives raise to a Green
function free of the on-shell $ \delta $ -function. In fact

\[ \alpha F + (1 - \alpha) \bar{F} = \alpha \left( W + i \pi \delta
\right) + \]
\[ (1- \alpha ) \left( W - i \pi \delta \right) = W + (2 \alpha - 1 )
i \pi \delta \]

 I.e.: only for $ \alpha $ = $ 1/2 $ the free propagation effects
disappear from the Green function.

\section{ Discussion }

 The negative outcome of the latest experiments on the Higgs,
suggests that at least the scalar sector of the standard model may
need some revision. However, in view of the succes of the model,
it seems convenient to reduce the changes to a minimun.

 From the point of view here adopted, the model is kept practically
untouched.

 The symmetry breaking mechanism is such that the lagrangian potential
for weak fields corresponds to a negative mass squared particle. It
is then possible to set up the hypotesis that the Higgs, just as the
tachyon, is unobservable as a free particle. In other words, it is
unable to occupy free asymptotic states.

The vacuum, defined as the zero energy eigenstate,
is now more symmetric with respect to the creation and annihilation
operators.

 The resultant propagator is half causal and half anticausal. It
is compatible with the adopted point of view. It is also the
unique combination of causal and anticausal Green functions that
eliminates the on-shell $ \delta $-function which represents the
free propagation of the particle. Furthermore, according to
ref. \cite{tp8}, in spite of the advanced components it contains,
it does not produce observable acausal effects.

 There are no differences in the predicted amplitudes when only
tree diagrams are involved. In those diagrams, if a virtual Higgs
is present, its propagator is off-shell and the $ \delta $ -function
has no influence. Feynman's and Wheeler's propagators give then
the same results. It is only through radiative corrections
containing a loop where a Higgs is present, that any difference
could be detected.

 In practice, our suggestion means that the amplitude for any
electroweak process is to be evaluated as usual. No Higgs will
ever appear in external legs. When a virtual Higgs appears in
any internal loop, a on-shell $ \delta $ -function contribution
must be sustracted (Cf. eq.(26)).

\pagebreak

\end{document}